\documentclass[twopage,11pt]{article}
\usepackage{times}
\usepackage{graphicx}
\usepackage{fancyhdr}
\usepackage{amssymb}
\usepackage{amsmath}
\usepackage{fancybox}

\begin{document}

\title{Topological Defects in Ferromagnetic, Antiferromagnetic and Cyclic
Spinor Condensates -- A Homotopy Theory}
\author{Yunbo Zhang$^{1,2}$, Harri M\"{a}kel\"{a}$^{1}$, and Kalle-Antti
Suominen$^{1}$ \\
$^{1}${\small Department of Physics, University of Turku, FIN-20014 Turun
yliopisto, Finland}\\
$^{2}${\small Department of Physics and Institute of Theoretical Physics, }\\
{\small Shanxi University, Taiyuan 03006, P. R. China}}
\date{\today}
\maketitle

\begin{abstract}
We apply the homotopy group theory in classifying the topological defects in
atomic spin-1 and spin-2 Bose-Einstein condensates. The nature of the
defects depends crucially on the spin-spin interaction between the atoms. We
find the topologically stable defects both for spin-1 ferromagnetic and
anti-ferromagnetic states, and for spin-2 ferromagnetic and cyclic states.
With this rigorous approach we clarify the previously controversial
identification of symmetry groups and order parameter spaces for the spin-1
anti-ferromagnetic state, and show that the spin-2 cyclic case provides a
rare example of a physical system with non-Abelian line defects, like those
observed in biaxial nematics. We also show the possibility to produce
vortices with fractional winding numbers of $1/2$, $1/3$ and their multiples
in spinor condensates.
\end{abstract}

\vspace*{-13cm} 
\begin{flushleft}
{\scriptsize To appear in \textit{
Progress in Ferromagnetic Research} \\ \vspace*{-0.1cm}
Frank Columbus, editor (Nova Science Publishers, New York, 2004)}
\end{flushleft}\vspace*{13cm}

\section{Introduction}

The all-optical trapping of Bose-Einstein condensates (BEC) \cite%
{Optical,Barrett} has opened up a new direction in the study of dilute
atomic gases, i.e., the spinor condensates with degenerate internal degrees
of freedom of the hyperfine spin $F$. For alkali atoms with $F=1$, both
experiments and theories have shown two possible kinds of spin correlations
in the atom species, namely ferromagnetic (e.g. $^{87}$Rb \cite%
{Ho,Ohmi,Chang}) or antiferromagnetic (e.g. $^{23}$Na \cite{Ho,Ohmi,Stenger}%
). With the experimental success of condensing alkali bosons with $F>1$ such
as $^{85}$Rb \cite{Cornish} and $^{133}$Cs~\cite{Weber}, and the unusual
stability of the $F=2$ state (against spin-exchange) in $^{87}$Rb~\cite%
{Suominen}, one expects that defects with much richer structure can be
created in the future. A remarkable feature here is that both the gauge
symmetry $\mathrm{U(1)}$ and the spin symmetry $\mathrm{SO(3)}$ are
involved, a situation similar to superfluid $^{3}$He where three different
continuous symmetries (orbital, spin and gauge) are broken either
independently or in a connected fashion~\cite{Helium,Salomaa}.

Topological defects and excitations in the spinor BECs have been studied
theoretically by several groups \cite%
{Ho,Ohmi,Khawaja,Stoof,Mizushima,Martikainen,Martikainen2,Yip,LeonhardtVolovik,Zhou1,Zhou2}%
. Stoof and Khawaja \cite{Khawaja} showed that ferromagnetic condensates
have long-lived Skyrmion excitations, which are nonsingular but
topologically nontrivial pointlike spin textures. Moreover, they also found
that spin-1 Bose-Einstein antiferromagnets have singular pointlike
topological spin textures \cite{Stoof}, which are analogous to the 't
Hooft-Polyakov magnetic monopoles in particle physics. Coreless vortices
were demonstrated to be thermodynamically stable in ferromagnetic $F=1$
spinor condensates under rotation \cite{Mizushima,Martikainen2} and were
phase imprinted in a $F=1$ sodium condensate experimentally \cite{Leanhardt}%
. Yip \cite{Yip} has performed a systematic study on vortex structures and
presented several axisymmetric and non-axisymmetric vortices for $F=1$
antiferromagnetic BEC. Martikainen et al. \cite{Martikainen} proposed and
demonstrated numerically a method to create monopoles in three dimensional
two-component condensates. Linear defects were studied by Leonhardt and
Volovik \cite{LeonhardtVolovik}, who pointed out the existence of Alice
strings in the condensate of $^{23}$Na.

Most of the work on this subject is based on the original identification of
the order parameter spaces by Ho~\cite{Ho}. After the original studies it
was also claimed by Zhou that a discrete symmetry of $Z_{2}$ type was missed
in the case of antiferromagnetic spin-1 condensate~\cite{Zhou1,Zhou2} and
therefore the topological defects would manifest totally different
structures. In this article we present a rigorous topological study that
both solves this spin-1 controversy, and reveals interesting aspects of
spin-2 systems. The phases of spin-2 spinor condensates are characterized by
a pair of parameters $\left\vert \left\langle \mathbf{F}\right\rangle
\right\vert $ and $|\Theta |$ describing the ferromagnetic order and the
formation of singlet pairs, respectively \cite%
{Ciobanu,Koashi,Ueda,MartikainenJPB}. It turns out that for the so called
cyclic phase the fundamental group that determines the nature of possible
stable topological defects is \textit{non-Abelian}. The only known physical
example of such a system so far has been the biaxial nematic liquid crystal.

The organization of this paper is as follows: In the following section, we
shall review the basic physics of the spinor condensate and discuss the
possible ground states for hyperfine spin $F=1$ and $F=2$. In Section 3, we
give a brief introduction of the homotopy theory of the defect
classification, taking the nematic liquid crystal and superfluid $^{3}$He as
examples. We present our calculation of the homotopy groups for spinor
condensates in Sections 4-7. The non-Abelian fundamental group for the
cyclic phase and its indications are discussed in detail and the order
parameter spaces are easily identified in a correct way following our
procedure of symmetry breaking. We summarize our results in Section 8.

\section{Spinor Condensate}

Neutral atomic gases can be confined in conventional magnetic traps with the
availability of hyperfine states being restricted by the requirement that
the trapped atoms remain in weak-field seeking states. Alkali atoms with a
nuclear spin of $I=3/2$, such as $^{87}$Rb and $^{23}$Na, have three
weak-field seeking states at small field. A far-off-resonant optical trap,
however, confines atoms regardless of their hyperfine state. Thus, the
atomic spin is liberated from the requirements of magnetic trapping and
becomes a new degree of freedom. In particular, all atoms in the lower
hyperfine manifold, for example the $F=1$ hyperfine manifold of sodium, can
be stably trapped simultaneously. Such multi-component optically trapped
condensates are represented by an order parameter which is a vector in
hyperfine spin space, and are thus called spinor Bose-Einstein condensates.
The spin relaxation collisions in spinor condensates allow for population
exchange among hyperfine states without trap loss. Theoretical studies
started with the determination of the ground state structure in mean field
theory for both spin-1 \cite{Ho,Ohmi} and spin-2 \cite%
{Ciobanu,Koashi,Ueda,MartikainenJPB,Klausen,HoYin} cases. Law et al. \cite%
{Law,Pu} investigated the spin correlation beyond mean-field limit and the
spin-mixing dynamics due to the nonlinear interaction in the spinor
condensate. The dynamics is sensitive to the relative phase and particle
number distribution among the individual components of the condensate. Ho
and Yip \cite{HoYip} later found that the ground state of a spin-1 Bose gas
with an antiferromagnetic interaction was a fragmented condensate in uniform
magnetic fields. Zhou \cite{Zhou1,Zhou2} showed that the low energy spin
dynamics in the system can be mapped into an $o(n)$ nonlinear sigma model.
The formation of ground state spin domains, metastable states and quantum
tunneling were observed in a series experiments at MIT \cite%
{Stenger,Miesner,Stamper-Kurn2,Stamper-Kurn}. The discussions in this paper,
however, mainly concern the possible ground states in mean-field theory.

\subsection{Spin-1 case}

The ground states of the spinor condensate are determined through the
minimization of the energy functional with the constraint of the
conservation of the atom number and magnetization \cite{Stamper-Kurn}. An $%
F=1$ spinor Bose-Einstein condensate is described by a three-component order
parameter $\psi (\mathbf{r})=\left\langle \hat{\Psi}(\mathbf{r}%
)\right\rangle =\left( \psi _{+1},\psi _{0},\psi _{-1}\right) ^{T}$. In
second quantized notation, the Hamiltonian describing a weakly-interacting
Bose gas can be obtained from the Gross-Pitaevskii theory \cite{Ho}%
\begin{eqnarray}
H &=&\int d^{3}\mathbf{r} {\Bigg \{}\hat{\Psi}_{i}^{\dag }(\mathbf{r})\left( -%
\frac{\hbar ^{2}\nabla ^{2}}{2m}+U(\mathbf{r})\right) \hat{\Psi}_{j}(\mathbf{%
r})\delta _{ij}  \notag \\
&&+\frac{1}{2}g_{0}\hat{\Psi}_{i}^{\dag }(\mathbf{r})\hat{\Psi}_{j}^{\dag }(%
\mathbf{r})\hat{\Psi}_{i}(\mathbf{r})\hat{\Psi}_{j}(\mathbf{r})  \notag \\
&&+\frac{1}{2}g_{2}\hat{\Psi}_{i}^{\dag }(\mathbf{r})\hat{\Psi}_{j}^{\dag }(%
\mathbf{r})\left( F_{a}\right) _{ik}\left( F_{a}\right) _{jl}\hat{\Psi}_{k}(%
\mathbf{r})\hat{\Psi}_{l}(\mathbf{r}) {\Bigg \}}
\end{eqnarray}%
where $\hat{\Psi}_{i}(\mathbf{r})$ is the field annihilation operator for an
atom with mass $m$ in hyperfine state $\left\vert 1,i\right\rangle $ at
position $\mathbf{r}$ with $i=+1,0,-1$ and $U(\mathbf{r})$ is the trapping
potential. Here the repeated indices are summed. The scattering lengths $%
a_{0}$ and $a_{2}$ characterize collisions between atoms through the total
spin $0$ and $2$ channels, respectively, $g_{0}=\frac{4\pi \hbar ^{2}}{m}%
\frac{a_{0}+2a_{2}}{3}$ is interaction strength through the
\textquotedblleft density\textquotedblright\ channel, and $g_{2}=\frac{4\pi
\hbar ^{2}}{m}\frac{a_{2}-a_{0}}{3}$ is that through the \textquotedblleft
spin\textquotedblright\ channel.

It is convenient to express the order parameter as $\psi (\mathbf{r})=\sqrt{%
n(\mathbf{r})}\zeta (\mathbf{r})$ where $n(\mathbf{r})$ is the atomic
density and $\zeta (\mathbf{r})$ is a three-component spinor $\zeta (\mathbf{%
r})=\left( \zeta _{+1},\zeta _{0},\zeta _{-1}\right) ^{T}=\left(
x_{+}e^{i\theta _{+}},x_{0}e^{i\theta _{0}},x_{-}e^{i\theta _{-}}\right)
^{T} $ of normalization $\left\vert \zeta \right\vert ^{2}=1$. Here $x$ and $%
\theta $ are the amplitudes and phases of the components. The spinor
determines the average local spin by means of $\left\langle \mathbf{F}%
\right\rangle =\zeta ^{\dag }(\mathbf{r})\mathbf{F}\zeta (\mathbf{r})$, and $%
\mathbf{F}$ are the usual spin-1 matrices with 
\begin{equation*}
F_{x}=\frac{1}{\sqrt{2}}\left( 
\begin{array}{ccc}
0 & 1 & 0 \\ 
1 & 0 & 1 \\ 
0 & 1 & 0%
\end{array}%
\right) ,F_{y}=\frac{i}{\sqrt{2}}\left( 
\begin{array}{ccc}
0 & -1 & 0 \\ 
1 & 0 & -1 \\ 
0 & 1 & 0%
\end{array}%
\right) ,F_{z}=\left( 
\begin{array}{ccc}
1 & 0 & 0 \\ 
0 & 0 & 0 \\ 
0 & 0 & -1%
\end{array}%
\right)
\end{equation*}%
which obey the commutation relations $\left[ F_{a},F_{b}\right] =i\epsilon
_{abc}F_{c}$. We thus obtain the energy functional%
\begin{eqnarray}
K &=&\int d^{3}\mathbf{r}\left\{ \psi ^{\dag }\left( -\frac{\hbar ^{2}\nabla
^{2}}{2m}\right) \psi +\left( U(\mathbf{r})-\mu \right) n+\frac{n^{2}}{2}%
\left( g_{0}+g_{2}\left\langle \mathbf{F}\right\rangle ^{2}\right) \right\} 
\notag \\
&=&\int d^{3}\mathbf{r}\left( K_{0}+n^{2}g_{2}\left\langle \mathbf{F}%
\right\rangle ^{2}/2\right)
\end{eqnarray}%
where $K_{0}$ is the density-dependent part and the chemical potential $\mu $
determines the number of atoms in the condensate. It is obvious that all
spinors related to each other by gauge transformation $e^{i\theta }$ and
spin rotations $\mathcal{U}=e^{-iF_{z}\alpha }e^{-iF_{y}\beta
}e^{-iF_{z}\gamma }$ are energetically degenerate in zero external magnetic
field, where $(\alpha ,\beta ,\gamma )$ are the Euler angles. The
ground-state spinor is determined by minimizing the spin-dependent
mean-field interaction energy, $n^{2}g_{2}\left\langle \mathbf{F}%
\right\rangle ^{2}/2$. There are two distinct states depending on the sign
of the interaction parameter $g_{2}$:

\begin{itemize}
\item $g_{2}>0$ (i.e. $a_{2}>a_{0}$, e.g. $^{23}$Na): anti-ferromagnetic or
polar state as the condensate lowers its energy by minimizing its average
spin, i.e. by making $\left\langle \mathbf{F}\right\rangle =0$. The ground
state spinor is then one of a degenerate set of spinors, the
\textquotedblleft polar\textquotedblright\ states, corresponding to all
possible rotations of the hyperfine state $m_{F}=0$, i.e.%
\begin{equation}
\zeta (\mathbf{r})=e^{i\theta }\mathcal{U}\left( 
\begin{array}{c}
0 \\ 
1 \\ 
0%
\end{array}%
\right) =e^{i\theta }\left( 
\begin{array}{c}
-\frac{1}{\sqrt{2}}e^{-i\alpha }\sin \beta \\ 
\cos \beta \\ 
\frac{1}{\sqrt{2}}e^{i\alpha }\sin \beta%
\end{array}%
\right)  \label{afm1}
\end{equation}

\item $g_{2}<0$ (i.e. $a_{2}<a_{0}$, e.g. $^{87}$Rb): ferromagnetic as the
condensate lowers its energy by maximizing its average spin, i.e. by making $%
\left\langle \mathbf{F}\right\rangle =1$. In this case the ground state
spinors correspond to all rotations of the hyperfine state $m_{F}=1$, i.e.%
\begin{equation}
\zeta (\mathbf{r})=e^{i\theta }\mathcal{U}\left( 
\begin{array}{c}
1 \\ 
0 \\ 
0%
\end{array}%
\right) =e^{i\left( \theta -\gamma \right) }\left( 
\begin{array}{c}
e^{-i\alpha }\cos ^{2}\frac{\beta }{2} \\ 
\sqrt{2}\cos \frac{\beta }{2}\sin \frac{\beta }{2} \\ 
e^{i\alpha }\sin ^{2}\frac{\beta }{2}%
\end{array}%
\right)  \label{fm1}
\end{equation}
\end{itemize}

\subsection{Magnetic Field}

One can tailor the ground state structure with an external magnetic field
and the effects of field inhomogeneities and quadratic Zeeman shifts modify
the spin-dependent interaction energy into \cite{Stenger}%
\begin{equation}
K_{spin}=\left( c\left\langle \mathbf{F}\right\rangle ^{2}-p\left\langle
F_{z}\right\rangle +q\left\langle F_{z}^{2}\right\rangle \right) n
\end{equation}%
where $c=g_{2}n/2$. The linear Zeeman shift $p=g\mu _{B}Bz+p_{0},$ where $g$
is the Land\'{e} $g$-factor and $\mu _{B}$ is the Bohr magneton, comes from
the field gradient $B$ along the long axis $z$ of the condensate, while the
last term gives the quadratic Zeeman shift from homogeneous field which is
always positive for spin-1 condensate in a weak field. Assuming conservation
of total spin, we have included a Lagrange multiplier $p_{0}$ into $p$. For
a system with zero total spin, $p_{0}$ cancels the linear Zeeman shift due
to a homogeneous bias $B_{0}$, yielding $p=0$. Positive (negative) values of 
$p$ are achieved for condensates with a positive (negative) overall spin.
The parameters $p$ and $q$ can be related to the individual level shifts by
(energies in units of the hyperfine splitting $E_{HFS}$) 
\begin{eqnarray}
2p &=&E_{-}-E_{+}  \notag \\
2q &=&E_{-}+E_{+}-2E_{0}
\end{eqnarray}%
where the Zeeman energies $E_{+},$ $E_{0}$ and $E_{-}$ of the $m_{F}=+1,0,-1$
can be expressed according to the Breit-Rabi formula \cite{Woodgate} as 
\begin{eqnarray}
E_{+} &=&-\frac{1}{8}-\frac{1}{2}\sqrt{1+x+x^{2}}  \notag \\
E_{0} &=&-\frac{1}{8}-\frac{1}{2}\sqrt{1+x^{2}}  \notag \\
E_{-} &=&-\frac{1}{8}-\frac{1}{2}\sqrt{1-x+x^{2}}
\end{eqnarray}%
with $x=g\mu _{B}B/E_{HFS}$.

Including the non-diagonal terms of the mean field interaction, we may
minimize the energy functional 
\begin{eqnarray}
K_{spin}/n &=&c\left( x_{+}^{2}-x_{-}^{2}\right) ^{2}+2cx_{0}^{2}\left(
x_{+}^{2}+x_{-}^{2}+2x_{+}x_{-}\cos \phi \right)  \notag \\
&&-p\left( x_{+}^{2}-x_{-}^{2}\right) +q\left( x_{+}^{2}+x_{-}^{2}\right)
\end{eqnarray}%
by means of the Lagrange multiplier method subjected to the constraint of
normalization 
\begin{equation}
g=x_{+}^{2}+x_{0}^{2}+x_{-}^{2}-1=0  \label{g}
\end{equation}%
where $\phi =\theta _{+}+\theta _{-}-2\theta _{0}.$ The solutions to the
first derivatives of the Lagrange multiplier function $X=K_{spin}/n-\lambda
g $ can be classified into the following table of spinors with their
corresponding energies

\begin{center}
\begin{tabular}{l|l|l}
& spinors & energies \\ \hline
1 & $e^{i\theta _{+1}}\left( 1,0,0\right) $ & $c-p+q$ \\ 
2 & $e^{i\theta _{-1}}\left( 0,0,1\right) $ & $c+p+q$ \\ 
3 & $e^{i\theta _{0}}\left( 0,1,0\right) $ & $0$ \\ 
4 & $\left( e^{i\theta _{+1}}\sqrt{\frac{2c+p}{4c}},0,e^{i\theta _{-1}}\sqrt{%
\frac{2c-p}{4c}}\right) $ & $q-\frac{p^{2}}{4c}$ \\ 
5 & $\left( e^{i\theta _{+1}}\sqrt{\frac{2c-p+q}{2c}},e^{i\theta _{0}}\sqrt{%
\frac{p-q}{2c}},0\right) $ & $\frac{\left( 2c-p+q\right) ^{2}}{4c}$ \\ 
6 & $\left( 0,e^{i\theta _{0}}\sqrt{\frac{-p-q}{2c}},e^{i\theta _{-1}}\sqrt{%
\frac{2c+p+q}{2c}}\right) $ & $\frac{\left( 2c+p+q\right) ^{2}}{4c}$ \\ 
7 & $\left( 
\begin{array}{c}
e^{i\theta _{+1}}\sqrt{\frac{q^{2}+4cq-p^{2}}{16cq^{3}}\left( q+p\right) ^{2}%
}, \\ 
e^{i\theta _{0}}\sqrt{\frac{-q^{2}+4cq-p^{2}}{8cq^{3}}\left(
q^{2}-p^{2}\right) }, \\ 
e^{i\theta _{-1}}\sqrt{\frac{q^{2}+4cq-p^{2}}{16cq^{3}}\left( q-p\right) ^{2}%
}%
\end{array}%
\right) $ & $\frac{\left( q^{2}+4cq-p^{2}\right) ^{2}}{16cq^{2}}$%
\end{tabular}
\end{center}

\medskip We notice that spinors 4-7 are only well-defined in some specific
regions in the $p$--$q$ plane, i.e., the quantities under the square root
must be non-negative. For example, spinor 7 may only exist for $%
q^{2}+4cq-p^{2}<0$ and $q^{2}>p^{2}$, and in addition we must have $\phi =0$
or $\pi $. The ground state spinors obtained by minimizing the energy
functional can be indicated in the so-called spin-domain phase diagrams
(Figure 1 in ref. \cite{Stenger}). For $c=0$, the Zeeman energy causes the
cloud to separate into three pure domains with $m_{F}=+1,0,-1$ and with
boundaries at $\left\vert p\right\vert =q$. For $c>0$, a spin domain with
mixed $m_{F}=\pm 1$ components, i.e., spinor 4, appears in the
anti-ferromagnetic phase diagram. For $c<0$, all three components are
generally miscible and have no sharp boundaries, which corresponds to spinor
7.

\subsection{Conservation of Magnetization}

Although conservation of the magnetization was included in the above
section, it was not separately discussed. Consequently the results do not
easily apply to systems with fixed values of the magnetization $\mathcal{M}$%
. The ground state structures as given in \cite{Stenger} correspond to the
actual ground state as realized through an $\mathcal{M}$ non-conserving
evaporation process (e.g. in the presence of a non-zero $B$-field) that
serves as a reservoir for condensate magnetization. On the other hand, the
phase diagram for fixed values of $\mathcal{M}$ was also explicitly
discussed \cite{ZhangWX}, which could physically correspond to experimental
ground states (with/without a $B$-field) due to an $\mathcal{M}$ conserving
evaporation process. This requires the introduction of two Lagrange
multipliers during the minimization subjected to conservation constraints
for both the atomic number $N$ and magnetization $\mathcal{M}$, which in the
mean-field approximation are given by 
\begin{eqnarray}
N &=&\int d^{3}\mathbf{r}n(\mathbf{r})\left( x_{+}^{2}(\mathbf{r})+x_{0}^{2}(%
\mathbf{r})+x_{-}^{2}(\mathbf{r})\right) ,  \notag \\
\mathcal{M} &=&\int d^{3}\mathbf{r}n(\mathbf{r})\left( x_{+}^{2}(\mathbf{r}%
)-x_{-}^{2}(\mathbf{r})\right) .
\end{eqnarray}
We restrict the discussion here to the situation that equation (\ref{g}) and 
\begin{equation}
h=x_{+}^{2}-x_{-}^{2}-m=0  \label{h}
\end{equation}%
are satisfied where $m=\mathcal{M}/N$. With the definition $%
x=x_{+}^{2}+x_{-}^{2}$, we can assort the possible spinors minimizing the
Lagrange multiplier function $X=K_{spin}/n-\lambda g-\delta h$ into the
following classes (where the energy zero point has been moved to $pm$):

\begin{center}
\begin{tabular}{l|l|l}
& spinors & energies \\ \hline
4 & $\left( e^{i\theta _{+1}}\sqrt{\frac{1+m}{2}},0,e^{i\theta _{-1}}\sqrt{%
\frac{1-m}{2}}\right) $ & $cm^{2}+q$ \\ 
5 & $\left( e^{i\theta _{+1}}\sqrt{m},e^{i\theta _{0}}\sqrt{1-m},0\right) $
& $-cm^{2}+\left( 2c+q\right) m$ \\ 
6 & $\left( 0,e^{i\theta _{0}}\sqrt{1+m},e^{i\theta _{-1}}\sqrt{-m}\right) $
& $-cm^{2}-\left( 2c+q\right) m$ \\ 
7 & $\left( e^{i\theta _{+1}}\sqrt{\frac{x_{m}+m}{2}},e^{i\theta _{0}}\sqrt{%
1-x_{m}},e^{i\theta _{-1}}\sqrt{\frac{x_{m}-m}{2}}\right) $ & $cm^{2}+g_{\pm
}(x_{m})+qx_{m}$%
\end{tabular}
\end{center}

\medskip We still have the spinors 1-3 which are the same as in above
section, however, they only exist for special values $m=+1,-1,0$,
respectively. While spinor 5(6) is confined to the positive (negative)
values of $m$, 4 and 7 may exist for the whole region $-1\leq m\leq 1$. In
spinor 7 with three nonzero components, the phase convention remains $\phi
=0 $ or $\pi $ and the minimum is reached when $x=x_{m}$ where $x_{m}$ is
determined by%
\begin{equation}
g_{\pm }^{\prime }(x_{m})+q=0
\end{equation}%
with $g_{\pm }(x)=2c(1-x)\left( x\pm \sqrt{x^{2}-m^{2}}\right) $ for
ferromagnetic($+$) or anti-ferromagnetic($-$) interaction, respectively. The
ground state spinor phase diagram for a homogeneous condensate may be
determined in the $m$--$q$ plane, as indicated for positive $m$ case in
Figure 4 of ref. \cite{ZhangWX}. For $c=0$, spinor 5 will always dominate
except that on the boundary $q=0$ we have spinor 7. For $c<0$, spinor 7 will
dominate. For $c>0$, a curve $q=2c(1-\sqrt{1-m^{2}})$ divides spinors 4 and
7.

\subsection{Spin-2 case}

For $^{23}$Na and $^{87}$Rb with regular hyperfine multiplets, the lower
hyperfine state $F=1$ has lower energy than the upper state $F=2$.
Experimentally only atoms in the lower hyperfine states can be confined in
the optical trap. Those in the upper hyperfine states will leave the trap by
spin-flip scattering. Since spin-flip scattering is strong in $^{23}$Na,
only the high-field seeking stretched state $\left\vert 2,-2\right\rangle $
exhibits reasonable stability, experiments with more complex spinor
condensate do not seem to be possible \cite{Gorlitz}. On the other hand,
optically trapped $^{87}$Rb has proved to be a candidate for spin-2 Bose gas 
\cite{Schmaljohann} with rich spin dynamics and magnetization conservation
was also observed during the mixing \cite{Chang}. In the case of $^{85}$Rb,
the lowest multiplet has spin $F=2$ and a negative $s$-wave scattering
length in zero field. With the success to Bose condense $^{85}$Rb in
magnetic traps \cite{Cornish}, it is conceivable that an $F=2$ spinor
condensate might be trapped optically in lower hyperfine states, provided
that the three particle losses when the field is reduced through the
Feshbach resonance are not too large.

Bose systems require that the total angular momentum of two colliding spin-2
particles is restricted to 0, 2, and 4. The effective low-energy Hamiltonian
including the interaction energy describing binary collisions via the $s$%
-wave scattering can be generally expressed as \cite{Ciobanu,Koashi,Ueda}%
\begin{eqnarray}
H &=&\int d^{3}\mathbf{r}{\Bigg \{}\hat{\Psi}_{i}^{\dag }(\mathbf{r})\left( -%
\frac{\hbar ^{2}\nabla ^{2}}{2m}+U(\mathbf{r})\right) \hat{\Psi}_{j}(\mathbf{%
r})\delta _{ij}  \notag \\
&&+\frac{1}{2}c_{0}\hat{\Psi}_{i}^{\dag }(\mathbf{r})\hat{\Psi}_{j}^{\dag }(%
\mathbf{r})\hat{\Psi}_{i}(\mathbf{r})\hat{\Psi}_{j}(\mathbf{r})  \notag \\
&&+\frac{1}{2}c_{1}\hat{\Psi}_{i}^{\dag }(\mathbf{r})\hat{\Psi}_{j}^{\dag }(%
\mathbf{r})\left( F_{a}\right) _{ik}\left( F_{a}\right) _{jl}\hat{\Psi}_{k}(%
\mathbf{r})\hat{\Psi}_{l}(\mathbf{r})  \notag \\
&&+\frac{1}{2}5c_{2}\hat{\Psi}_{i}^{\dag }(\mathbf{r})\hat{\Psi}_{j}^{\dag }(%
\mathbf{r})\left\langle 2i;2j|00\right\rangle \left\langle
00|2k;2l\right\rangle \hat{\Psi}_{k}(\mathbf{r})\hat{\Psi}_{l}(\mathbf{r})%
{\Bigg \}}
\end{eqnarray}%
where $\left\langle 00|2k;2l\right\rangle $ is the Clebsch-Gordan
coefficient for combining two spin-2 particles with $m_{F}=k$ and $l$ into a
spin singlet $\left\vert 0,0\right\rangle $. The parameters 
\begin{eqnarray}
c_{0} &=&\frac{4\pi \hbar ^{2}}{m}\frac{4a_{2}+3a_{4}}{7},  \notag \\
c_{1} &=&\frac{4\pi \hbar ^{2}}{m}\frac{a_{4}-a_{2}}{7},  \notag \\
5c_{2} &=&\frac{4\pi \hbar ^{2}}{m}\frac{3a_{4}-10a_{2}+7a_{0}}{7}
\end{eqnarray}%
describe the density-density interaction, spin-spin interaction, and
formation of the singlet pair, respectively. The spinor $\zeta (\mathbf{r})$
with five components $\zeta (\mathbf{r})=\left( \zeta _{+2},\zeta
_{+1},\zeta _{0},\zeta _{-1},\zeta _{-2}\right) $ normalized to unity,
determines the average local spin as $\left\langle \mathbf{F}\right\rangle
=\zeta ^{\dag }(\mathbf{r})\mathbf{F}\zeta (\mathbf{r})$, and $\mathbf{F}$
are the $5\times 5$ spin-2 matrices which obey the same commutation
relations $\left[ F_{a},F_{b}\right] =i\epsilon _{abc}F_{c}$. In the
mean-field approach the properties of a spinor condensate are determined by
the spin-dependent energy functional%
\begin{equation}
K_{spin}=\left( c_{1}\left\langle \mathbf{F}\right\rangle
^{2}+c_{2}\left\vert \Theta \right\vert ^{2}-p\left\langle
F_{z}\right\rangle +q\left\langle F_{z}^{2}\right\rangle \right) n
\label{k2}
\end{equation}%
where $\Theta =2\zeta _{+2}\zeta _{-2}-\zeta _{+1}\zeta _{-1}+\zeta _{0}^{2}$
represents a singlet pair of identical spin-2 particles and is invariant
under any rotation. The parameters $p$ and $q$ are related to the individual
level shifts by%
\begin{eqnarray}
p &=&\frac{1}{12}\left( E_{+2}-E_{-2}\right) +\frac{2}{3}\left(
E_{-1}-E_{+1}\right)  \notag \\
q &=&-\frac{1}{24}\left( E_{+2}+E_{-2}\right) +\frac{2}{3}\left(
E_{-1}+E_{+1}\right) -\frac{5}{4}E_{0}
\end{eqnarray}%
The Breit-Rabi formula \cite{Woodgate} in the case of $^{23}$Na or $^{87}$Rb
($F=2$ is the upper hyperfine state with higher energy) gives 
\begin{eqnarray*}
E_{+2} &=&-\frac{1}{8}+\frac{1}{2}\left( 1+x\right) \\
E_{+1} &=&-\frac{1}{8}+\frac{1}{2}\sqrt{1+x+x^{2}} \\
E_{0} &=&-\frac{1}{8}+\frac{1}{2}\sqrt{1+x^{2}} \\
E_{-1} &=&-\frac{1}{8}+\frac{1}{2}\sqrt{1-x+x^{2}} \\
E_{-2} &=&-\frac{1}{8}+\frac{1}{2}\left( 1-x\right)
\end{eqnarray*}%
In weak field, the quadratic Zeeman splitting is always negative, i.e., $q=-%
\frac{1}{16}x^{2}+O\left( x^{4}\right) $. In the case of $^{85}$Rb ($I=5/2$)
the lowest multiplet has spin $F=2$. From the individual level shift 
\begin{eqnarray*}
E_{+2} &=&-\frac{1}{12}-\frac{1}{2}\sqrt{1+\frac{4}{3}x+x^{2}} \\
E_{+1} &=&-\frac{1}{12}-\frac{1}{2}\sqrt{1+\frac{2}{3}x+x^{2}} \\
E_{0} &=&-\frac{1}{12}-\frac{1}{2}\sqrt{1+x^{2}} \\
E_{-1} &=&-\frac{1}{12}-\frac{1}{2}\sqrt{1-\frac{2}{3}x+x^{2}} \\
E_{-2} &=&-\frac{1}{12}-\frac{1}{2}\sqrt{1-\frac{4}{3}x+x^{2}}
\end{eqnarray*}%
$\allowbreak $we easily see $q$ is always positive for $^{85}$Rb at small
field, $q=\allowbreak \frac{1}{36}x^{2}+O\left( x^{4}\right) $. Unlike in
the case of spin-1, the whole $p$--$q$ plane is accessible experimentally
for a spin-2 condensate.

The ground state magnetization must be aligned with the external field, i.e.
along $z$-axis, implying $\left\langle \mathbf{F}\right\rangle
^{2}=\left\langle F_{z}\right\rangle ^{2}$ in eq. (\ref{k2}). Minimization
of the spin dependent energy functional using the similar Lagrange
multiplier method leads to three possible phases, one more compared to the
spin-1 case. These phases are characterized by a pair of parameters $%
\left\vert \left\langle \mathbf{F}\right\rangle \right\vert $ and $%
\left\vert \Theta \right\vert $ describing the ferromagnetic order and the
formation of singlet pairs, respectively. For convenience, we only consider
the linear Zeeman shift\ $p$ due to the magnetic field:

\begin{itemize}
\item Polar/Anti-ferromagnetic phases%
\begin{eqnarray*}
P &:&\sqrt{\frac{1}{2}}\left( e^{i\theta _{+2}}\sqrt{1+\frac{p}{4c_{1}-c_{2}}%
},0,0,0,e^{i\theta _{-2}}\sqrt{1-\frac{p}{4c_{1}-c_{2}}}\right) \\
P_{1} &:&\sqrt{\frac{1}{2}}\left( 0,e^{i\theta _{+1}}\sqrt{1+\frac{p}{%
2(c_{1}-c_{2})}},0,e^{i\theta _{-1}}\sqrt{1-\frac{p}{2(c_{1}-c_{2})}}%
,0\right) \\
P_{0} &:&e^{i\theta _{0}}\left( 0,0,1,0,0\right)
\end{eqnarray*}%
with energies $c_{2}-p^{2}/\left( 4c_{1}-c_{2}\right) ,c_{2}-p^{2}/4\left(
c_{1}-c_{2}\right) $ and $c_{2}$ respectively. Here $\theta _{i}$ are
arbitrary phases for the corresponding components. These states are
energetically degenerate in the absence of the external field with energy $%
c_{2}$ and parameters $\left\langle \mathbf{F}\right\rangle =0$ and $%
\left\vert \Theta \right\vert =1$.

\item Ferromagnetic phases%
\begin{eqnarray*}
F &:&e^{i\theta _{+2}}\left( 1,0,0,0,0\right) \\
F^{\prime } &:&e^{i\theta _{+1}}\left( 0,1,0,0,0\right)
\end{eqnarray*}%
with energies $4c_{1}-2p$ and $c_{1}-p$ respectively. This phase has a
non-vanishing parameter $\left\vert \left\langle \mathbf{F}\right\rangle
\right\vert =1$ indicating the ferromagnetic order and $\left\vert \Theta
\right\vert =0$.

\item Cyclic phase%
\begin{equation*}
C:\frac{1}{2}\left( e^{i\phi }\left( 1+\frac{p}{4c_{1}}\right) ,0,\sqrt{2-%
\frac{p^{2}}{8c_{1}^{2}}},0,e^{-i\phi }\left( -1+\frac{p}{4c_{1}}\right)
\right)
\end{equation*}%
with energy $-p^{2}/4c_{1}$ and $\phi$ an arbitrary phase. This is a
nonmagnetic phase which has no spin-1 analog and was referred to as the
cyclic state because of its close analog to the $d$-wave BCS\ superfluids.
Both parameters are zero, $\left\langle \mathbf{F}\right\rangle =0$ and $%
\left\vert \Theta \right\vert =0$.
\end{itemize}

Recent experiments observed clear evidence of polar behaviour for $F=2$
spinor condensate of $^{87}$Rb, and the slow dynamics of prepared cyclic
ground states showed the $F=2$ state to be close to the cyclic phase \cite%
{Klausen,Schmaljohann}. However, the nature of the spinor condensate which
depends on the $s$-wave scattering lengths for the total spins 0, 2, and 4,
may be changed into other phases by an offset magnetic field.

\section{Outline of the homotopy theory of defects}

We sketch out the procedure which has been widely used in the study of
topological defects in ordered media such as liquid crystals, superfluid $%
^{3}$He and heavy-fermion superconductors. The explicit use of homotopy for
topological classification of defects was made by some French \cite%
{Kleman,Michel,Toulouse,Poenaru}\ and Russian authors \cite{Volovik,Mineev}.
The results were well summarized in two review articles \cite{Michel,Mermin}%
. The central feature of the classification scheme of the defects emerges
from examining the mappings of closed curves in physical space into the
order-parameter space (OPS).

The order parameter of a system has associated with it a group of
transformations $G$. The set of all transformations in $G$ that leave the
reference order parameter $f$ (chosen arbitrarily but thereafter fixed)
unchanged is known as the isotropy group $H=\{g\in G|gf=f\}$. The OPS can
then be taken to be the space of cosets of $H$ in $G$: $M=G/H$. In terms of
broken symmetry, the fact that the ordering breaks the underlying symmetry
is expressed in the fact that $H$ is only a subgroup of $G$. The description
that follows will be valid for any group $G$ that \textit{acts transitively}
on $M$, i.e., if $f_{1}$ and $f_{2}$ are possible values of the order
parameter, then there is a transformation $g$ in $G$ which takes $f_{1}$
into $f_{2}$: $f_{2}=gf_{1}$.

Homotopy groups of the order-parameter space describe physical defects \cite%
{Mermin}. The $n$-th homotopy group $\pi _{n}(M)$ of the space $M$ consists
of the equivalence classes of continuous maps from $n$-dimensional sphere $%
S_{n}$ to the space $M$. Two maps are equivalent if they are homotopic to
one another. In three dimensional space, the first homotopy group, also
called the fundamental group, $\pi _{1}(M)$ describes singular line defects
and domain walls, which are non-singular defects. The second homotopy group $%
\pi _{2}(M)$ describes singular point defects and non-singular line defects.
\ These can be calculated with the help of the fundamental theorem: Let $G$
be a \textit{connected}, \textit{simply connected} continuous group and $%
H_{0}$ be the set of points in $H$ that can be connected to the identity by
a continuous path lying entirely in $H$. Then we have the isomorphisms 
\begin{equation}
\pi _{1}(M)=H/H_{0},\qquad \pi _{2}(M)=\pi _{1}(H_{0}).
\end{equation}%
For the theorem to hold, it is necessary that $\pi _{0}(G)=\pi _{1}(G)=\pi
_{2}(G)=0,$ meaning that $G$ has only one connected piece, any loop in $G$
can be shrunk continuously to a point, and $G$ has a vanishing second
homotopy group. While the second homotopy groups are always Abelian, the
fundamental groups can either be Abelian (each element constitutes a
conjugacy class), or non-Abelian (the line defects are characterized by the
conjugacy classes instead of the elements). In Figure 1 we give a schematic
description of the procedure for calculation of homotopy groups.

\begin{center}
\includegraphics[width=5in]{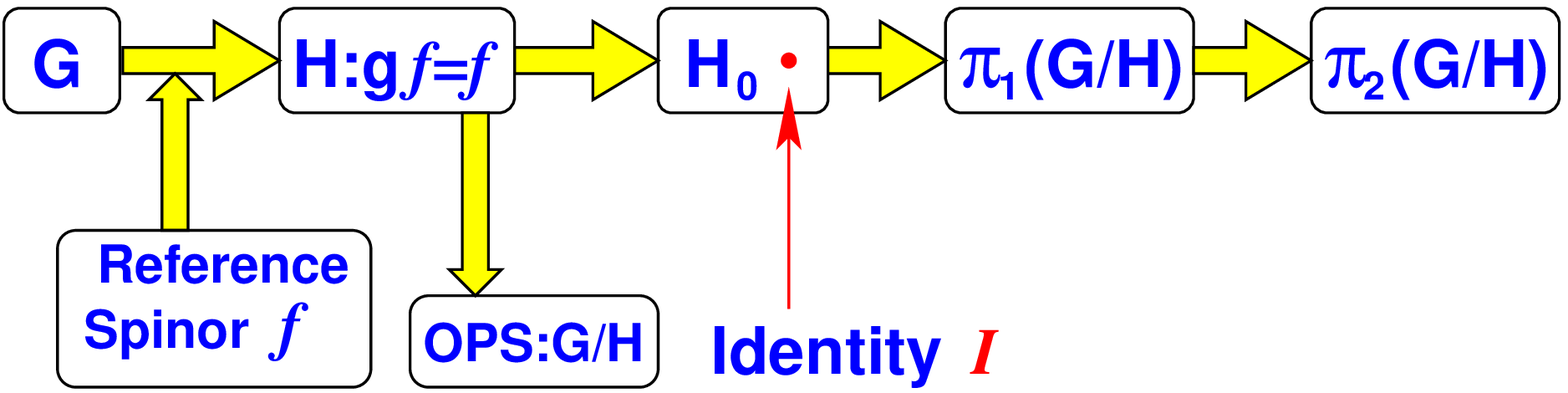}
\\
\bf{Figure 1: A schematic description of the procedure for calculation of
homotopy groups.}
\end{center}

A ready example for illustrating the above procedure is the biaxial
nematics, whose symmetry is that of a rectangular box (proper point group $%
D_{2}$). If $G$ is taken to be $\mathrm{SO(3)}$ then the isotropy subgroup $%
H $ is the four-element group consisting of the identity and $180^{\circ }$
rotations about three mutually perpendicular axes ($D_{2}$). Order parameter
space is thus identified as $M=\mathrm{SO(3)}/D_{2}$. If, however, we take $%
G $ to be $\mathrm{SU(2)}$, the universal covering group of $\mathrm{SO(3)}$%
, then $H$ is expanded to the non-Abelian quaternion group $Q$ (known as the 
\textit{lift} or \textit{double group}) with eight elements%
\begin{equation}
Q=\{\pm 1,\pm i\sigma _{x},\pm i\sigma _{y},\pm i\sigma _{z}\}.
\end{equation}%
The natural representation for the order parameter space of a biaxial
nematic turns out to be $M=\mathrm{SU(2)}/Q$. Since it is a discrete
subgroup of \textrm{SU(2)}, $H/H_{0}=H$. Thus $\pi _{1}(M)=Q$, and $\pi
_{2}(M)=0$. There are no stable point defects in biaxial nematics and the
line defects are characterized by five conjugacy classes of group $Q$%
\begin{eqnarray}
C_{0} &=&\{1\},\overline{C_{0}}=\{-1\},  \notag \\
C_{x} &=&\{\pm i\sigma _{x}\},C_{y}=\{\pm i\sigma _{y}\},C_{z}=\{\pm i\sigma
_{z}\}.
\end{eqnarray}%
The class $C_{0}$ contains removable trivial defects; $\overline{C_{0}}$
contains defects in which the object rotates about 360$^{\circ }$ as the
defect line is encircled; the other three classes contain defects in which
the rotation is through 180$^{\circ }$ about each of the three distinct
symmetry axes. The defects here are non-commutative, providing an example
with non-Abelian fundamental group.

Another illustrative example is the dipole-free $A$-phase of $^{3}$He, which
affords an unusual example of a case where $G$ must be bigger than $\mathrm{%
SO(3)}$. The order parameter is the product of an arbitrary unit 3-vector $%
\hat{n}$ and a complex 3-vector of the form $\hat{u}+i\hat{v}$, where $\hat{u%
}$ and $\hat{v}$ are an orthonormal pair. The orientations of $\hat{n}$ and $%
\hat{u}+i\hat{v}$ are uncoupled. Take the reference order parameter to be $%
A_{ij}=z_{i}(x_{j}+iy_{j})$, the group $G$ can be taken to be the direct
product of $\mathrm{SO(3)}$ with itself: $G=\mathrm{SO(3)\times SO(3)}$,
elements of $G$ consisting of pairs $(R,R^{\prime })$ of distinct rotations.
The isotropy group $H$ consists of elements of the form $\left( R(\hat{z}%
,\theta ),1\right) $ and $\left( R(\hat{u},\pi ),R(\hat{z},\pi )\right) $
for any axis $\hat{u}$ in the $x$--$y$ plane. To construct a simply
connected $G$, we must replace each $\mathrm{SO(3)}$ by $\mathrm{SU(2)}$.
Determining the \textit{lift} of $H$ from $\mathrm{SO(3)\times SO(3)}$ to
its covering group $\mathrm{SU(2)\times SU(2)}$, we find the isotropy group
consists of 4 pieces $\left\{ H_{0},gH_{0,}g^{2}H_{0},g^{3}H_{0}\right\} $
with the connected component of the identity $H_{0}=\left\{ (u(\hat{z}%
,\theta ),1)\right\} $ and $g=\left( u\left( \hat{x},\pi \right) ,u\left( 
\hat{z},\pi \right) \right) $. In this article, the notations $R$ and $u$
represent the rotations in $\mathrm{SO(3)}$ and $\mathrm{SU(2)}$,
respectively. The fundamental group is thus isomorphic to the cyclic group
of order 4, $\pi _{1}(M)=Z_{4}$ and $\pi _{2}(M)=Z$.

For the spinor condensate it seems natural to identify the underlying
symmetry group as $\mathrm{U(1)\times SO(3)}$, the groups in the direct
product representing the gauge and spin degrees of freedom respectively.
This group is \textit{not} simply connected, i.e., $\pi _{1}(\mathrm{%
U(1)\times SO(3)})\neq 0$. To apply the theorem, however, it is again
essential that one chooses the group $G$ to be simply connected. We proceed
by specifying the symmetry group as its universal covering group $R\times 
\mathrm{SU(2)}$, with the group of real numbers $R$ representing any
translation $\theta \in (-\infty ,+\infty )$ in the phase of the condensate.
For $F=1$, we use the 3D representation of the group $\mathrm{SU(2)}$ in
order to obtain the isotropy group, e.g., a rotation $u(z,\alpha )$ around
axis $z$ by angle $\alpha $ takes the form of a diagonal matrix $Diag\left(
e^{-i\alpha },1,e^{i\alpha }\right) $, a rotation $u(y,\beta )$ around axis $%
y$ by angle $\beta $ takes the form of%
\begin{equation*}
\left( 
\begin{array}{ccc}
\frac{1}{2}\left( 1+\cos \beta \right) & \frac{-\sin \beta }{\sqrt{2}} & 
\frac{1}{2}\left( 1-\cos \beta \right) \\ 
\frac{\sin \beta }{\sqrt{2}} & \cos \beta & \frac{-\sin \beta }{\sqrt{2}} \\ 
\frac{1}{2}\left( 1-\cos \beta \right) & \frac{\sin \beta }{\sqrt{2}} & 
\frac{1}{2}\left( 1+\cos \beta \right)%
\end{array}%
\right) .
\end{equation*}%
The two elements $\pm u(z,\alpha )$ are represented by the same matrix $%
Diag\left( e^{-i\alpha },1,e^{i\alpha }\right) $ in this even representation
of $\mathrm{SU(2)}$, though we know (and should always bear in mind) that $%
u(z,\alpha +2\pi )=-u(z,\alpha )$ while $u(z,\alpha +4\pi )=u(z,\alpha )$.

\section{Calculation of the homotopy groups}

There are two possible ground states in $F=1$ case. For the ferromagnetic
state, the isotropy group $H$ is constructed by the set of transformations
which leave the reference order parameter $(1,0,0)^{T}$ invariant. From the
degenerate family of the ground state spinor eq.(\ref{fm1}) we know
immediately that the angles should satisfy 
\begin{equation}
\beta =0,\theta -\alpha -\gamma =2n\pi
\end{equation}%
with $n$ an integer. The elements in group $H$ are the combination of a
translational part and a rotational part $H=\left\{ \left( \theta
,u(z,\theta )\right) ,\left( \theta ,u(z,\theta +2\pi )\right) \right\}
=\left\{ \left( \theta ,\pm u(z,\theta )\right) \right\} $. Evidently this
group includes two disconnected components---the connected component of the
identity $H_{0}=\left\{ \left( \theta ,u(z,\theta )\right) \right\} $ is
isomorphic to $R$. The group $H/H_{0}$ is isomorphic to the integers modulo
2, i.e., $Z_{2}$. The second homotopy group $\pi _{2}$ is trivial and we
arrive at the same result as that in Ref. \cite{Khawaja} 
\begin{equation}
\pi _{1}(M)=Z_{2},\qquad \pi _{2}(M)=0\text{, (spin-1 FM state). }
\end{equation}%
A ferromagnetic spin-1 condensate may have therefore only singular vortices
with winding number one while the point-like defects are topologically
unstable. Alternatively we may take the symmetry group as $\mathrm{SU(2)}$
because we can produce all possible gauge transformations by absorbing $%
\theta $ into the Euler angle $\gamma $. The isotropy group is discrete and
isomorphic to $Z_{2}$, which gives exactly the same result.

The polar state emerges if the atoms in the condensate interact
anti-ferromagnetically. In the ground state eq. (\ref{afm1}), the reference
parameter $(0,1,0)^{T}$ is left invariant for just those elements with 
\begin{equation}
\beta =0,\theta =2n\pi \quad \text{ or }\quad \beta =\pi ,\theta =\left(
2n+1\right) \pi .  \label{p1}
\end{equation}%
Thus the isotropy group $H$ includes now the transformations in which both
the rotation and the translation leave the spinor unchanged, and those in
which the rotation takes the reference spinor $(0,1,0)^{T}$ to $(0,-1,0)^{T}$
and the translation takes it back, i.e., a $\pi $ rotation about arbitrary
axis perpendicular to $\hat{z}$ combined with a $\pi $ translation in $%
\theta $ (or any odd multiples of $\pi $). The latter invariance is
identical to the Ising gauge symmetry emphasized in eq. (14) of Ref. \cite%
{Zhou2}. The full isotropy group is the union of these two sets, $H=\left\{
\left( 2n\pi ,u(z,\alpha )\right) ,\left( (2n+1)\pi ,gu(z,\alpha )\right)
\right\} $ where $g=u(y,\pi )$. There are infinitely many discrete
components in $H$, while the connected component of the identity $%
H_{0}=\left\{ \left( 0,u(z,\alpha )\right) \right\} $ is isomorphic to $%
\mathrm{U(1)}$. The elements with an even translational parity are of the
form $(2n\pi ,I)H_{0}$, and those with an odd parity are of the form $\left(
(2n+1)\pi ,g\right) H_{0}$. The group $H/H_{0}$ is therefore isomorphic to
the group of integers $Z$ through the isomorphism $\left( (2n+j)\pi
,g^{j}\right) H_{0}\mapsto 2n+j$ for $j=0,1$. We recover the conclusion that
line and point defects in spin-1 polar state can be classified by integer
winding numbers, 
\begin{equation}
\pi _{1}(M)=Z,\qquad \pi _{2}(M)=Z\text{, (spin-1 Polar state).}
\end{equation}%
Thus the $Z_{2}$ term does not appear in the homotopy group. We argue that
the identification of the OPS in Ref.~\cite{Zhou1,Zhou2} is also incorrect
(see below). Physically there are indeed infinite number of line defects
corresponding to integer and half-integer vortices (eq. (27) in Ref. \cite%
{Zhou2}). On the other hand, it is the Ising symmetry that leads to
half-vortices ($j=1$), which have been shown to be the unique linear defects
in polar condensate in addition to the usual integer vortices ($j=0$) \cite%
{LeonhardtVolovik}. If we move around a closed path in the condensate we
note that when we return to the starting point the angle $\theta $ has
changed by some amount. If we define the change in this angle divided by $%
2\pi $ to be the winding number, we see from the elements of $H/H_{0}$ that
the winding number can be either an integer $n$ or a half-integer $n+1/2$.

\section{Spin-2 Bose condensate}

We next apply the same approach to the BEC of spin-2 bosons. The defects
which may be created in spin-2 condensate exhibit even more elaborate
structures due to quantum correlations among bosons. For $F=2$ we have to
use the 5D representation of $SU(2)$, e.g., the rotation $u(z,\alpha )$ is
represented by matrix $Diag\left( e^{-2i\alpha },e^{-i\alpha },1,e^{i\alpha
},e^{2i\alpha }\right) $ and $u(y,\beta )$ takes the form of%
\begin{equation*}
\left( 
\begin{array}{ccccc}
\cos ^{4}\frac{\beta }{2} & -\sin \beta \cos ^{2}\frac{\beta }{2} & \frac{%
\sqrt{6}}{4}\sin ^{2}\beta & -\sin \beta \sin ^{2}\frac{\beta }{2} & \sin
^{4}\frac{\beta }{2} \\ 
\sin \beta \cos ^{2}\frac{\beta }{2} & \frac{\cos \beta +\cos 2\beta }{2} & -%
\frac{\sqrt{6}}{4}\sin 2\beta & \frac{\cos \beta -\cos 2\beta }{2} & -\sin
\beta \sin ^{2}\frac{\beta }{2} \\ 
\frac{\sqrt{6}}{4}\sin ^{2}\beta & \frac{\sqrt{6}}{4}\sin 2\beta & \frac{%
1+3\cos 2\beta }{4} & -\frac{\sqrt{6}}{4}\sin 2\beta & \frac{\sqrt{6}}{4}%
\sin ^{2}\beta \\ 
\sin \beta \sin ^{2}\frac{\beta }{2} & \frac{\cos \beta -\cos 2\beta }{2} & 
\frac{\sqrt{6}}{4}\sin 2\beta & \frac{\cos \beta +\cos 2\beta }{2} & -\sin
\beta \cos ^{2}\frac{\beta }{2} \\ 
\sin ^{4}\frac{\beta }{2} & \sin \beta \sin ^{2}\frac{\beta }{2} & \frac{%
\sqrt{6}}{4}\sin ^{2}\beta & \sin \beta \cos ^{2}\frac{\beta }{2} & \cos ^{4}%
\frac{\beta }{2}%
\end{array}%
\right)
\end{equation*}%
The calculations of the degenerate family of the ground state spinors and
the corresponding homotopy groups are straightforward and some results have
been reported in \cite{Harri}. Here we pick up some interesting features in
our results, focusing on the symmetry properties of the defects in
comparison with those in other ordered media. We first consider the defects
in the absence of an external field and the effect of magnetic field will be
discussed later.

We start with the case of the ferromagnetic state $F$. Equating the general
expression for the ground state spinor 
\begin{equation}
\zeta =e^{i(\theta -2\gamma )}\left( 
\begin{array}{c}
e^{-2i\alpha }\allowbreak \cos ^{4}\frac{\beta }{2} \\ 
e^{-i\alpha }\sin \beta \cos ^{2}\frac{\beta }{2} \\ 
\frac{\sqrt{6}}{4}\sin ^{2}\beta \\ 
e^{i\alpha }\sin \beta \sin ^{2}\frac{\beta }{2} \\ 
e^{2i\alpha }\sin ^{4}\frac{\beta }{2}%
\end{array}%
\right)  \label{f}
\end{equation}%
with the reference spinor $(1,0,0,0,0)^{T}$ leads to the requirement for the
isotropy group $H$ 
\begin{equation}
\beta =0,\theta -2\alpha -2\gamma =2n\pi .
\end{equation}%
We see that taking $n=0,1,2,3$ is enough for all possible transformations,
with the translational part being arbitrary and the rotational part
containing the rotations around axis $z$ by $\theta /2,\theta /2+\pi ,\theta
/2+2\pi ,\theta /2+3\pi $ respectively. Hence the group $H$ is composed of
four pieces $H=\left\{ \left( \theta ,u(z,n\pi +\theta /2)\right) \right\} $%
. Here it is important to show that the four components are not connected:
there does not exist a continuous path in $H$ which connects one component
to another, though the rotational parts themselves are connected. The
connected component of the identity $H_{0}=\left\{ \left( \theta ,u(z,\theta
/2)\right) \right\} $ is again isomorphic to $R$. If we define an element $g$
of the group $R\times \mathrm{SU(2)}$ by $\left( 0,u(z,\pi )\right) $, we
see that the quotient group $H/H_{0}$ has the same structure as the cyclic
group of order 4, i.e., $\{e,g,g^{2},g^{3}\}$ and we conclude that 
\begin{equation}
\pi _{1}(M)=Z_{4},\qquad \pi _{2}(M)=0\text{, (spin-2 }F\text{ state).}
\label{fm2}
\end{equation}

It is interesting to check how the group $Z_{4}$ characterizes vortices for
state $F$. In spin-1 case there is only one topologically stable line
defect, that is, a vortex with winding number one. Equation (\ref{fm2})
shows that there are three stable vortices for spin-2 condensates. We can
set $\theta -2\gamma =2m\varphi $, $-\alpha =m\varphi $, $\beta =\pi t$ in
the ground state for $F$ state, Eq. (\ref{f}), which leads to a family of
spinor states parametrized by a parameter $t$ between 0 and 1. Here $m>0$ is
an integer, $\varphi $ is the azimuthal angle. When $t$ evolves from 0 to 1,
the $4m\varphi $ vortex state $\zeta (t=0)=\left( e^{i4m\varphi
},0,0,0,0\right) ^{T}$ evolves continuously to the vortex free state $\zeta
(t=1)=\left( 0,0,0,0,1\right) ^{T}$. This shows that vortices with winding
number $4m$ are topologically unstable. Similarly, by multiplying factors $%
e^{ik\varphi }(k=1,2,3)$ one obtains the following correspondences 
\begin{equation}
e^{i\left( 4m+k\right) \varphi }\left( 1,0,0,0,0\right) ^{T}\rightarrow
e^{ik\varphi }\left( 0,0,0,0,1\right) ^{T}
\end{equation}%
i.e., the vortices with winding numbers $4m+k$ may evolve into vortices with
winding numbers $k$, respectively. There are thus three classes of
topologically stable line defects. Together with the uniform state, they
form the fundamental group $Z_{4}$. Non-trivial vortices are those in which
the reference spinor rotates through 180$^{\circ }$, 360$^{\circ }$ or 540$%
^{\circ }$ about the $z$-axis when the defect line is circulated.
Straightforwardly for ferromagnetic condensates with spin $F,$ the
fundamental group $\pi _{1}(M)=Z_{2F}$ characterizes $(2F-1)$ classes of
stable line defects.

Spin variations in the ferromagnetic states in general lead to superflows 
\cite{Ho,MartikainenJPB}. To illustrate the coreless (or Skyrmion) vortices
in spin-2 case, we set $\theta -2\tau =2\varphi ,\alpha =\varphi $ in the
spinor degenerate family (\ref{f}) and consider the condensate 
\begin{equation}
\zeta (\mathbf{r})=\left( 
\begin{array}{c}
\allowbreak \cos ^{4}\frac{\beta }{2} \\ 
e^{i\varphi }\sin \beta \cos ^{2}\frac{\beta }{2} \\ 
e^{i2\varphi }\frac{\sqrt{6}}{4}\sin ^{2}\beta \\ 
e^{i3\varphi }\sin \beta \sin ^{2}\frac{\beta }{2} \\ 
e^{i4\varphi }\sin ^{4}\frac{\beta }{2}%
\end{array}%
\right)
\end{equation}%
where $\beta =\beta (r)$ is an increasing function of $r$ starting from $%
\beta =0$ at $r=0.$The superfluid velocity does not depend on $z$ and it is
cylindrically symmetric 
\begin{equation}
\mathbf{v}_{s}=\frac{\hbar }{M}\left[ 2\nabla \varphi -2\cos \beta \nabla
\varphi \right] =\frac{2\hbar }{Mr}\left( 1-\cos \beta \right) \hat{\varphi}
\end{equation}%
i.e., the coreless vortex may exist in the spin-2 case, with only the
velocity doubling its value compared to the spin-1 case \cite{Ho}. The
velocity vanishes instead of diverging at $r=0$ because $\beta (0)=0$. This
is called a coreless vortex. For a Mermin-Ho vortex \cite{MerminHo}, the
bending angle $\beta $ must be $\pi /2$ at the boundary of the condensate,
while for an Anderson-Toulouse \cite{AndersonToulouse} vortex $\beta $ must
be $\pi ,$ i.e. 
\begin{eqnarray}
\beta (R) &=&\pi /2,\text{ for Mermin-Ho}  \notag \\
\beta (R) &=&\pi ,\text{ for Anderson-Toulouse.}
\end{eqnarray}

\section{Non-Abelian homotopy groups}

Media with non-Abelian fundamental groups are especially interesting from
the topological point of view. The only illustrative example in ordered
media so far have been biaxial nematic liquid crystals~\cite{Yu}. Their
multiplication table has been verified experimentally \cite{DeNeve}.

We have found that the cyclic state $C$ provides another physically
realistic example in which the fundamental group is non-commutative. A
rotation and a gauge transformation of the reference spinor $\frac{1}{2}%
\left( e^{i\phi },0,\sqrt{2},0,-e^{-i\phi }\right) ^{T}$ in zero field
produce the following degenerate family 
\begin{equation*}
\zeta =\frac{1}{2}e^{i\theta }\left( 
\begin{array}{c}
e^{-2i\alpha }\left( \cos ^{4}\frac{\beta }{2}e^{i\phi -2i\tau }+\frac{\sqrt{%
3}}{2}\sin ^{2}\beta -\sin ^{4}\frac{\beta }{2}e^{-i\phi +2i\tau }\right) \\ 
e^{-i\alpha }\sin \beta \left( \cos ^{2}\frac{\beta }{2}e^{i\phi -2i\tau }-%
\sqrt{3}\cos \beta +\sin ^{2}\frac{\beta }{2}e^{-i\phi +2i\tau }\right) \\ 
\frac{\sqrt{6}}{4}\sin ^{2}\beta e^{i\phi -2i\tau }+\frac{\sqrt{2}}{4}%
(1+3\cos 2\beta )-\frac{\sqrt{6}}{4}\sin ^{2}\beta e^{-i\phi +2i\tau } \\ 
e^{i\alpha }\sin \beta \left( \sin ^{2}\frac{\beta }{2}e^{i\phi -2i\tau }+%
\sqrt{3}\cos \beta +\cos ^{2}\frac{\beta }{2}e^{-i\phi +2i\tau }\right) \\ 
e^{2i\alpha }\left( \sin ^{4}\frac{\beta }{2}e^{i\phi -2i\tau }+\frac{\sqrt{3%
}}{2}\sin ^{2}\beta -\cos ^{4}\frac{\beta }{2}e^{-i\phi +2i\tau }\right)%
\end{array}%
\right)
\end{equation*}
The reference spinor is left invariant by the elements of three sets
characterized by the translations in the phase of the condensate $\theta :$

\begin{itemize}
\item For $\theta =2n\pi ,$ one must have $\beta =0,\alpha +\tau =m\pi ,$ or 
$\beta =\pi ,\alpha -\tau =-\phi +\frac{\pi }{2}+m\pi ;$

\item For $\theta =\frac{2\pi }{3}+2n\pi ,$ one must have $\beta =\frac{\pi 
}{2},\alpha +\tau =-\frac{\pi }{2}+m\pi ,\alpha -\tau =-\phi +m^{\prime }\pi
;$

\item For $\theta =\frac{4\pi }{3}+2n\pi ,$ one must have $\beta =\frac{\pi 
}{2},\alpha +\tau =\frac{\pi }{2}+m\pi ,\alpha -\tau =-\phi +m^{\prime }\pi
. $
\end{itemize}

Here $m$ and $m^{\prime }$ are integers satisfying $m+m^{\prime }=odd$. For
all possible transformations we need take the values $m,m^{\prime }=0,1,2,3$
so that there are eight possibilities 
\begin{eqnarray*}
m &=&0,m^{\prime }=1,3 \\
m &=&1,m^{\prime }=0,2 \\
m &=&2,m^{\prime }=1,3 \\
m &=&3,m^{\prime }=0,2
\end{eqnarray*}%
This gives the isotropy group 
\begin{eqnarray}
H &=&\{\pm I,\pm a,\pm b,\pm c,  \label{c} \\
&&\pm d,\pm e,\pm f,\pm g,  \notag \\
&&\pm d^{2},\pm e^{2},\pm f^{2},\pm g^{2}\}.  \notag
\end{eqnarray}%
The spin rotations $a=u(z,\pi )$, $b=u(y,\pi )u(z,\phi +\pi /2)$ and $c=ba$
satisfy $a^{2}=b^{2}=c^{2}=-I$, while $d=u(z,\pi /4+\phi /2)u(y,\pi
/2)u(z,\pi /4-\phi /2)$, $e=-da$, $f=-ad$ and $g=-ada$ satisfy $%
d^{3}=e^{3}=f^{3}=g^{3}=-I$. Each element in the first, second, third row is
associated with an additional phase change $2n\pi ,2\pi /3+2n\pi ,4\pi
/3+2n\pi $ respectively. It is a discrete group, and $H_{0}$ consists of the
identity $\left( 0,I\right) $ only. The fundamental theorems identify that 
\begin{equation}
\pi _{1}(M)=H,\qquad \pi _{2}(M)=0\text{, (spin-2 }C\text{ state).}
\label{p2}
\end{equation}%
The elements in the fundamental group are non-commutative, for example $%
ab=-c\neq ba$.

The criterion for the topological equivalence of defects applies in the most
general case in terms of conjugacy classes of the fundamental group. Two
line defects are topologically equivalent if and only if they are
characterized by the same conjugacy class. Defects can still be labelled by
the elements of the first homotopy group, but if these elements belong to
the same conjugacy class, corresponding defects can be continuously
transformed to one another. However, if they belong to different conjugacy
classes this is not possible. It is thus necessary to classify the group
into the following conjugacy classes: 
\begin{eqnarray*}
C_{0}(n) &=&\{I\}_{n},\qquad \overline{C_{0}}(n)=\{-I\}_{n},\qquad
C_{2}(n)=\{\pm a,\pm b,\pm c\}_{n}, \\
C_{3}(n+1/3) &=&\{d,e,f,g\}_{n+1/3},\qquad \overline{C_{3}}%
(n+1/3)=\{-d,-e,-f,-g\}_{n+1/3}, \\
C_{3}^{2}(n+2/3) &=&\{d^{2},e^{2},f^{2},g^{2}\}_{n+2/3},\quad \overline{%
C_{3}^{2}}(n+2/3)=\{-d^{2},-e^{2},-f^{2},-g^{2}\}_{n+2/3}
\end{eqnarray*}%
with the subscripts standing for the winding numbers of the defects.
Physically this indicates the feasibility of creating not only vortices with
any integer winding number but also fractional quantum vortices. The class $%
C_{0}(n)$ describes defects in which the phase of the spinor is changed by $%
2\pi n$ as the defect line is encircled. Note that only $C_{0}(0)$
corresponds to trivial defects. In the case of $\overline{C_{0}}(n)$ phase
change of $2\pi n$ is accompanied by a 360$^{\circ }$ rotation about $z$%
-axis. The element $a$ with winding number $n$ in the class $C_{2}(n)$
depicts a defect in which the spinor rotates through 180$^{\circ }$ about
the $z$-axis and changes phase by $2\pi n$ as the line is encircled. The
multiplication table of conjugacy classes is shown in table 1. Only half of
table is shown because the class multiplication is commutative. Winding
numbers have been omitted for clarity. When two classes are multiplied the
winding number of the resulting class is the sum of the individual winding
numbers. It shows that, for example, when we combine defect $C_{2}(n)$ with $%
C_{2}(-n)$ they can either annihilate each other ($C_{0}(0)$) or form defect 
$\overline{C_{0}}(0)$ or $C_{2}(0)$, the result depending on how they are
brought together. Interesting features of this non-Abelian fundamental group
include the topological instability of the defects and their interaction,
i.e., entanglement when two of them are brought to cross with each other 
\cite{Mermin}.

\begin{center}
\bigskip Table 1: The multiplication table of the conjugacy classes of the
cyclic phase.

\medskip 
\begin{tabular}{lllllll}
\hline
& $\overline{C_{0}}$ & $C_{2}$ & $C_{3}$ & $\overline{C_{3}}$ & $C_{3}^{2}$
& $\overline{C_{3}^{2}}$ \\ \hline\hline
$\overline{C_{0}}$ & $C_{0}$ &  &  &  &  &  \\ 
$C_{2}$ & $C_{2}$ & $6C_{0}+6\overline{C_{0}}+4C_{2}$ &  &  &  &  \\ 
$C_{3}$ & $\overline{C_{3}}$ & $3\left( C_{3}+\overline{C_{3}}\right) $ & $3%
\overline{C_{3}^{2}}+C_{3}^{2}$ &  &  &  \\ 
$\overline{C_{3}}$ & $C_{3}$ & $3\left( C_{3}+\overline{C_{3}}\right) $ & $%
\overline{C_{3}^{2}}+3C_{3}^{2}$ & $3\overline{C_{3}^{2}}+C_{3}^{2}$ &  & 
\\ 
$C_{3}^{2}$ & $\overline{C_{3}^{2}}$ & $3\left( C_{3}^{2}+\overline{C_{3}^{2}%
}\right) $ & $4\overline{C_{0}}+2C_{2}$ & $4C_{0}+2C_{2}$ & $3C_{3}+%
\overline{C_{3}}$ &  \\ 
$\overline{C_{3}^{2}}$ & $C_{3}^{2}$ & $3\left( C_{3}^{2}+\overline{C_{3}^{2}%
}\right) $ & $4C_{0}+2C_{2}$ & $4\overline{C_{0}}+2C_{2}$ & $C_{3}+3%
\overline{C_{3}}$ & $3C_{3}+\overline{C_{3}}$ \\ \hline
\end{tabular}
\end{center}

The defects can be further grouped into classes, which form an Abelian group
isomorphic to the first homology group of the order parameter space \cite%
{Trebin}. This coarser classification is more general than the homotopic one
because two defects are considered equivalent also, if they can be
transformed into each other via a \textit{catalyzation} process consisting
of splitting a line singularity into two and recombining them beyond a third
one. All elements labelled by elements of the commutator subgroup $D$ of $%
\pi _{1}(M)$ can be catalyzed away by this procedure. $D$ is generated by
the commutators $\delta \tau \delta ^{-1}\tau ^{-1}$ of all pairs of
elements $\delta ,\tau \in \pi _{1}(M)$. The elements of $\pi _{1}(M)/D$ are
unions of conjugacy classes. In our case $D$ is the union of the conjugacy
classes with winding number zero, $D=C_{0}(0)\cup \overline{C_{0}}(0)\cup
C_{2}(0)$ and the first homology group is 
\begin{equation}
\pi _{1}(M)/D=\left\{ C_{0}\cup \overline{C_{0}}\cup C_{2},C_{3}\cup 
\overline{C_{3}},C_{3}^{2}\cup \overline{C_{3}^{2}}\right\}
\end{equation}%
The homology theory assembles the conjugacy classes further into three sets
for each $n$, in which the defects are labeled by the winding numbers $%
n,n+1/3,n+2/3$ respectively \cite{Harri}. Two defects in the same conjugacy
classes can be continuously converted into one another by local surgery,
while two defects in the same homology class can be deformed into one
another by the \textit{catalyzation} procedure.

\section{Order Parameter Spaces}

Like the quaternion group $Q$ for biaxial nematics, the fundamental group (%
\ref{c}) is the \textit{lift} of a point group in $R\times \mathrm{SU(2)}$.
To find the remaining discrete symmetry group for the cyclic state, and, in
addition, to clarify the controversial identification of the OPS for spin-1
case, in the remaining of this paper we turn to describe the system in terms
of rotations in $\mathrm{SO(3)}$, e.g., two elements $\pm u(z,\alpha )$ in $%
\mathrm{SU(2)}$ are mapped into one $R(z,\alpha )$ in $\mathrm{SO(3)}$ with $%
R(z,\alpha +2\pi )=R(z,\alpha )$.

The OPS for $F=1$ polar state was identified as $\mathrm{U(1)}\times S^{2}$
in Refs. \cite{Ho,Khawaja}. An extra $Z_{2}$ symmetry was claimed in Ref. 
\cite{Zhou1} so the author concluded the OPS as $\mathrm{U(1)}\times
S^{2}/Z_{2}$. Here we show that previous studies are incorrect. Taking the
group $G$ as $\mathrm{U(1)}_{G}\times \mathrm{SO(3)}_{S}$ where the
subscripts stand for the gauge and spin symmetries respectively, we see that
the isotropy group $H$ consists of two separate parts, $\left\{ \left(
e^{i0},R(z,\alpha )\right) \right\} $ and $\left\{ \left( e^{i\pi },R(y,\pi
)R(z,\alpha )\right) \right\} $. The rotations in the first part constitute
the group $\mathrm{SO(2)}$, while the elements in the second part are just
those in the group $\mathrm{O(2)}$ but not in $\mathrm{SO(2)}$ with
determinants $-1$. The combination of these two parts gives the full
isotropy group as $\mathrm{O(2)}$ where both gauge and spin symmetries are
involved. The OPS is the quotient $G/H=\left( \mathrm{U(1)}_{G}\times 
\mathrm{SO(3)}_{S}\right) /\mathrm{O(2)}_{G+S}$ and here it is not possible
to apply the fundamental theorem for $G$ is not any more simply connected.
One may wonder if we can factorize the OPS further as 
\begin{eqnarray*}
G/H &=&\left( \mathrm{U(1)}_{G}\times \mathrm{SO(3)}_{S}\right) /\mathrm{O(2)%
}_{G+S} \\
&=&\left( \mathrm{U(1)}_{G}\times \mathrm{SO(3)}_{S}\right) /(\mathrm{%
SO(2)\times Z}_{2})_{G+S} \\
&=&\mathrm{U(1)}\times S^{2}/Z_{2}
\end{eqnarray*}%
However it is incorrect because though in 3 dimensional space we have $%
\mathrm{O(3)}=\mathrm{SO(3)\times Z}_{2}$ but it is not true in 2
dimensional case, i.e. $\mathrm{O(2)}\neq \mathrm{SO(2)\times Z}_{2}$. The
spin and gauge symmetries are broken in a connected fashion just as in the
system of $^{3}$He \cite{Helium,Salomaa}. Table 2 summarizes our result in
comparison with the previous studies.\medskip

\begin{center}
Table 2: Comparison of the OPS and fundamental groups for spin-1 polar
condensate

\medskip 
\begin{tabular}{l|ll}
\hline
& OPS & $\pi _{1}(M)$ \\ \hline
Ho, Stoof, etc. & $\mathrm{U(1)}\times S^{2}$ & $Z$ \\ 
Zhou & $\mathrm{U(1)}\times S^{2}/Z_{2}$ & $Z\times Z_{2}$ \\ 
This paper & $\left( \mathrm{U(1)}\times \mathrm{SO(3)}\right) /\mathrm{O(2)}
$ & $Z$ \\ \hline
\end{tabular}%
\end{center}

\begin{center}
\includegraphics[width=4.0 in]{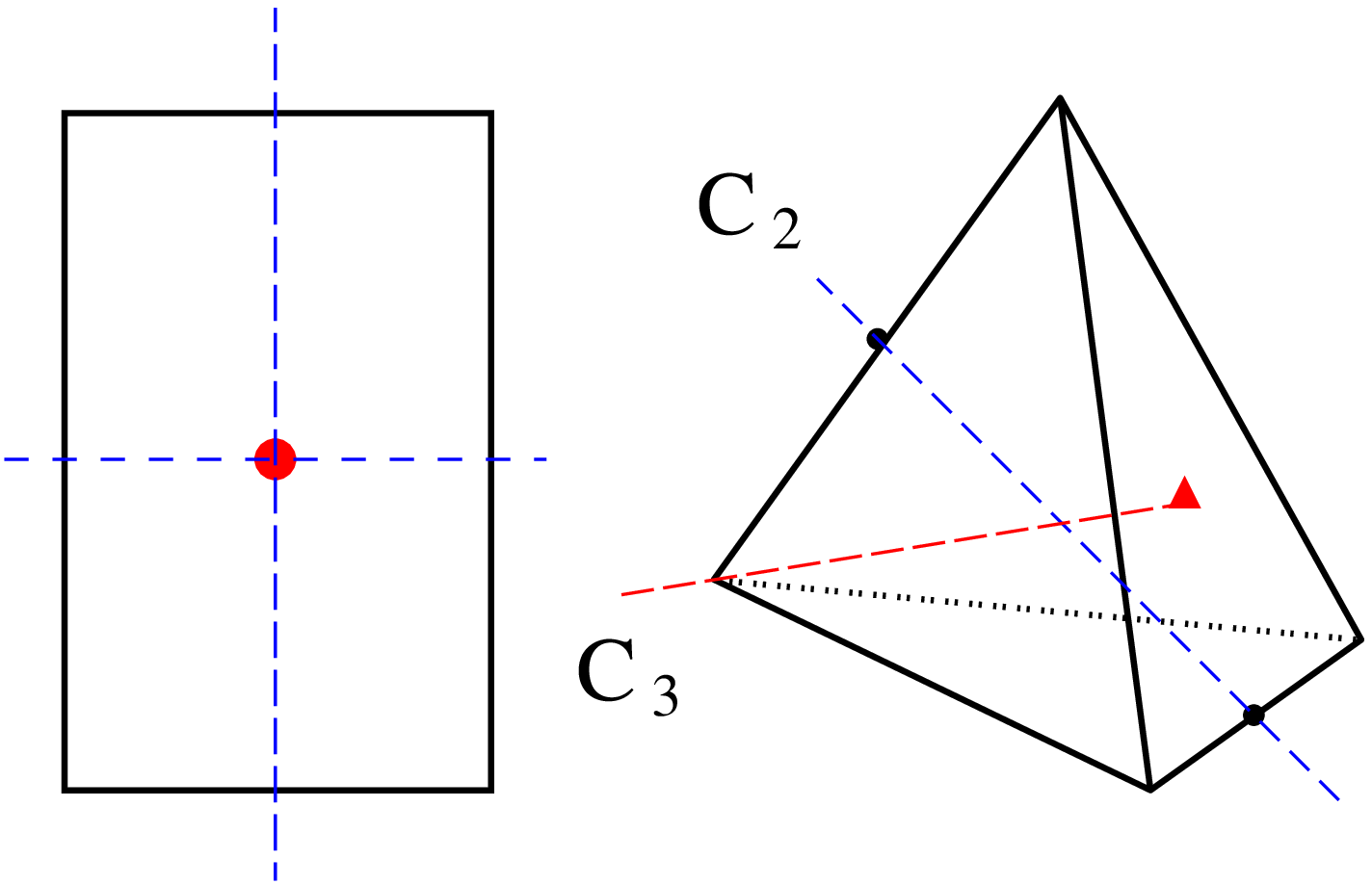}
\\
\bf{Figure 2: Symmetries of the defects in biaxial nematics ($D_{2}$) and cyclic
state $C$ in spin-2 condensate ($T$). The dot at the center of the rectangle
stands for axis $z$. The dashed lines represent 2-fold axes, except that
with a triangle for 3-fold axis.}
\end{center}

For the ferromagnetic state the group $H$ may be obtained if one notices
that the $2\pi $ difference in the rotational angle does not give another
component as it did in the case of $\mathrm{SU(2)}$. We have $H=\left\{
\left( e^{i\theta },R(z,\theta )\right) \right\} $ which is isomorphic to $%
\mathrm{U(1)}_{G+S}.$ This means that there is a remaining symmetry $\mathrm{%
U(1)}$ in the symmetry broken system. The OPS is thus factorized as $\left( 
\mathrm{U(1)}_{G}\times \mathrm{SO(3)}_{S}\right) /\mathrm{U(1)}_{G+S}=%
\mathrm{SO(3)}_{S+G}$.

The discrete symmetry group of defects in the spin-2 cyclic state $C$ can be
shown to be isomorphic to the tetrahedral group $T$. We continue to
represent $G$ as $\mathrm{U(1)}_{G}\times \mathrm{SO(3)}_{S}$. The isotropy
group (\ref{c}) is shrunk to a group of 12 elements if one understands the
rotation in the sense of $\mathrm{SO(3)}$ (i.e., $a=R(z,\pi )$), 
\begin{equation}
H=\{I,a,b,c,\varepsilon d,\varepsilon e,\varepsilon f,\varepsilon
g,\varepsilon ^{2}d^{2},\varepsilon ^{2}e^{2},\varepsilon
^{2}f^{2},\varepsilon ^{2}g^{2}\},  \label{cc}
\end{equation}%
where $\varepsilon =\exp (2\pi i/3)$ comes from the gauge transformation and 
$\varepsilon d$, for instance, is an abbreviation for the element $\left(
\varepsilon ,d\right) $. Three 2-fold rotational axes are $z$ and 2 lines in 
$xy$ plane perpendicular to each other (which lie on axes $x$ and $y$ if we
choose the arbitrary phase $\phi =\pi /2$). The elements $\varepsilon
d,\varepsilon e,\varepsilon f,\varepsilon g$ are four 3-fold axes. The
symmetries remaining in the symmetry broken states for biaxial nematics and
spin-2 cyclic state are shown in Figure 2. The OPS for state $C$ can be
identified as $\left( \mathrm{U(1)}_{G}\times \mathrm{SO(3)}_{S}\right)
/T_{G+S}$.

It should be noted that an applied magnetic field $B$ changes the defect
structure severely by reducing the degenerate family of the spinor. We take
again the cyclic state as an example. The symmetry group in this case is $%
\mathrm{U(1)\times SO(2)}$ because the magnetic field chooses its direction
automatically as the quantization axis. From the spinor 
\begin{equation*}
\frac{1}{2}\left( 
\begin{array}{c}
(1+p)e^{i\phi _{1}} \\ 
0 \\ 
\sqrt{2-2p^{2}}e^{i(\phi _{1}+\phi _{2})/2} \\ 
0 \\ 
(-1+p)e^{i\phi _{2}}%
\end{array}%
\right) 
\end{equation*}%
where $\phi _{1,2}$ are two arbitrary phases and $p\sim B$, we easily see
the possibility to create vortices in any of the three nonzero components
with winding number for $i-$th component $n_{i}$ confined by $%
n_{1}+n_{5}=2n_{3}$.

\section{Summary}

Our main findings are summarized in Table 3. We have determined the nature
of the topological defects in spin-1 and spin-2 condensates. The order
parameter spaces are identified as the spaces of the coset of the isotropy
group $H$ in the transformation group $G$. Topologically stable vortices
with winding numbers larger than unity may be created in the ferromagnetic
state for condensates with $F>1$, up to the value $(2F-1)$. The line defects
in the spin-2 cyclic state $C$ exhibit non-commutative features, resulting
e.g. in line defects with winding numbers of $1/3$ and its multiples. It
also turns out that in the zero field $\mathrm{U(1)\times SO(3)}$ does not
act transitively on the order-parameter space of the polar phase and thus
the defect structure remains unsolved.\medskip

\begin{center}
\medskip Table 3: Main results on calculation of the OPS and homotopy groups

\medskip 
\begin{tabular}{l|lll}
\hline
& OPS & $\pi _{1}$ & $\pi _{2}$ \\ \hline
Spin-1 FM & $\mathrm{SO(3)}$ & $Z_{2}$ & $0$ \\ 
Spin-1 AFM & $\left( \mathrm{U(1)}\times \mathrm{SO(3)}\right) /\mathrm{O(2)}
$ & $Z$ & $Z$ \\ 
Spin-2 $F$ & $\mathrm{SO(3)}/Z_{2}$ & $Z_{4}$ & $0$ \\ 
Spin-2 $F^{\prime }$ & $\mathrm{SO(3)}$ & $Z_{2}$ & $0$ \\ 
Spin-2 $C$ & $\left( \mathrm{U(1)}\times \mathrm{SO(3)}\right) /T$ & $H$ eq.(%
\ref{c}) & $0$ \\ \hline
\end{tabular}
\end{center}

\medskip \textbf{Acknowledgements:} The authors acknowledge the Academy of
Finland (Project No 206108) for financial support. YZ is also supported by
NSF of China (Grant Nos. 10175039 and 90203007). Discussions with O. K\"{a}%
rki and J. Hietarinta are appreciated.

\end{document}